\documentclass[aps,prl,preprint,superscriptaddress,showpacs]{revtex4}

\usepackage{graphicx}

\begin{document}


\title{Backward-angle $\eta$ photoproduction from 
protons at $E_\gamma$ = 1.6 -- 2.4 GeV}






\author{M.~Sumihama}
  \affiliation{Research Center for Nuclear Physics, Osaka University, Ibaraki, Osaka 567-0047, Japan}
\author{D.S.~Ahn}
  \affiliation{Research Center for Nuclear Physics, Osaka University, Ibaraki, Osaka 567-0047, Japan}
  \affiliation{Department of Physics, Pusan National University, Busan 609-735, Korea}
\author{J.K.~Ahn}
  \affiliation{Department of Physics, Pusan National University, Busan 609-735, Korea}
\author{H.~Akimune}
  \affiliation{Department of Physics, Konan University, Kobe, Hyogo 658-8501, Japan}
\author{Y.~Asano}
  \affiliation{Japan Synchrotron Radiation Research Institute, Sayo, Hyogo 679-5198, Japan}
\author{W.C.~Chang}
  \affiliation{Institute of Physics, Academia Sinica, Taipei 11529, Taiwan}
\author{S.~Dat\'{e}}
  \affiliation{Japan Synchrotron Radiation Research Institute, Sayo, Hyogo 679-5198, Japan}
\author{H.~Ejiri}
  \affiliation{Research Center for Nuclear Physics, Osaka University, Ibaraki, Osaka 567-0047, Japan}
\author{H.~Fujimura}
  \affiliation{Laboratory of Nuclear Science, Tohoku University, Sendai, Miyagi 982-0826, Japan}
\author{M.~Fujiwara}
  \affiliation{Research Center for Nuclear Physics, Osaka University, Ibaraki, Osaka 567-0047, Japan}
\author{S.~Fukui}
  \affiliation{Department of Physics and Astrophysics, Nagoya University, Aichi 464-8602, Japan}
\author{S.~Hasegawa}
  \affiliation{J-PARC Center, Japan Atomic Energy Agency, Tokai-mura, Ibaraki, 
319-1195, Japan}
\author{K.~Hicks}
  \affiliation{Department of Physics and Astronomy, Ohio University, Athens, Ohio 45701, USA}
\author{T.~Hotta}
  \affiliation{Research Center for Nuclear Physics, Osaka University, Ibaraki, Osaka 567-0047, Japan}
\author{K.~Imai}
  \affiliation{Department of Physics, Kyoto University, Kyoto 606-8502, Japan}
\author{T.~Ishikawa}
  \affiliation{Laboratory of Nuclear Science, Tohoku University, Sendai, Miyagi 982-0826, Japan}
\author{T.~Iwata}
  \affiliation{Department of Physics, Yamagata University, Yamagata 990-8560, Japan}
\author{Y.~Kato}
  \affiliation{Research Center for Nuclear Physics, Osaka University, Ibaraki, Osaka 567-0047, Japan}
\author{H.~Kawai}
  \affiliation{Department of Physics, Chiba University, Chiba 263-8522, Japan}
\author{K.~Kino}
  \affiliation{Graduate School of Engineering, Hokkaido University, Sapporo, Hokkaido 060-8628, Japan}
\author{H.~Kohri}
  \affiliation{Research Center for Nuclear Physics, Osaka University, Ibaraki, Osaka 567-0047, Japan}
\author{N.~Kumagai}
  \affiliation{XFEL Project Head Office, RIKEN, Sayo, Hyogo 679-5198, Japan}
\author{T.~Matsuda}
  \affiliation{Department of Applied Physics, Miyazaki University, Miyazaki 889-2192, Japan}
\author{T.~Matsumura}
  \affiliation{National Defense Academy in Japan, Yokosuka, Kanagawa 239-8686, Japan}
\author{T.~Mibe}
  \affiliation{High Energy Accelerator Research Organization, KEK, Tsukuba, Ibaraki, 305-0801, Japan}
\author{M.~Miyabe}
  \affiliation{Research Center for Nuclear Physics, Osaka University, Ibaraki, Osaka 567-0047, Japan}
\author{N.~Muramatsu}
  \affiliation{Research Center for Nuclear Physics, Osaka University, Ibaraki, Osaka 567-0047, Japan}
\author{T.~Nakano}
  \affiliation{Research Center for Nuclear Physics, Osaka University, Ibaraki, Osaka 567-0047, Japan}
\author{M.~Niiyama}
  \affiliation{RIKEN, The Institute of Physical and Chemical Research, Wako, Saitama 351-0198, Japan}
\author{M.~Nomachi}
  \affiliation{Department of Physics, Osaka University, Toyonaka, Osaka 560-0043, Japan}
\author{Y.~Ohashi}
  \affiliation{Japan Synchrotron Radiation Research Institute, Sayo, Hyogo 679-5198, Japan}
\author{H.~Ohkuma}
  \affiliation{Japan Synchrotron Radiation Research Institute, Sayo, Hyogo 679-5198, Japan}
\author{T.~Ooba}
  \affiliation{Department of Physics, Chiba University, Chiba 263-8522, Japan}
\author{D.S.~Oshuev}
  \affiliation{Institute of Physics, Academia Sinica, Taipei 11529, Taiwan}
\author{C.~Rangacharyulu}
  \affiliation{Department of Physics and Engineering Physics, University of Saskatchewan, Saskatoon SK
S7N 5E2, Canada}
\author{A.~Sakaguchi}
  \affiliation{Department of Physics, Osaka University, Toyonaka, Osaka 560-0043, Japan}
\author{P.M.~Shagin}
  \affiliation{School of Physics and Astronomy, University of Minnesota, Minneapolis, Minnesota 55455, USA}
\author{Y.~Shiino}
  \affiliation{Department of Physics, Chiba University, Chiba 263-8522, Japan}
\author{A.~Shimizu}
  \affiliation{Research Center for Nuclear Physics, Osaka University, Ibaraki, Osaka 567-0047, Japan}
\author{H.~Shimizu}
  \affiliation{Laboratory of Nuclear Science, Tohoku University, Sendai, Miyagi 982-0826, Japan}
\author{Y.~Sugaya}
  \affiliation{Department of Physics, Osaka University, Toyonaka, Osaka 560-0043, Japan}
\author{Y.~Toi}
  \affiliation{Department of Applied Physics, Miyazaki University, Miyazaki 889-2192, Japan}
\author{H.~Toyokawa}
  \affiliation{Japan Synchrotron Radiation Research Institute, Sayo, Hyogo 679-5198, Japan}
\author{A.~Wakai}
  \affiliation{Akita Research Institute of Brain and Blood Vessels, Akita, 010-0874, Japan}
\author{C.W.~Wang}
  \affiliation{Institute of Physics, Academia Sinica, Taipei 11529, Taiwan}
\author{S.C.~Wang}
  \affiliation{Institute of Physics, Academia Sinica, Taipei 11529, Taiwan}
\author{K.~Yonehara}
  \affiliation{Illinois Institute of Technology, Chicago, Illinois 60616, USA}
\author{T.~Yorita}
  \affiliation{Research Center for Nuclear Physics, Osaka University, Ibaraki, Osaka 567-0047, Japan}
\author{M.~Yoshimura}
  \affiliation{Institute for Protein Research, Osaka University, Suita, Osaka, 565-0871, Japan}
\author{M.~Yosoi}
  \affiliation{Research Center for Nuclear Physics, Osaka University, Ibaraki, Osaka 567-0047, Japan}
\author{R.G.T.~Zegers}
  \affiliation{National Superconducting Cyclotron Laboratory, Michigan State University, Michigan 48824-1321, USA}
\collaboration{LEPS Collaboration}
\noaffiliation
\date{\today}

\begin{abstract}
Differential cross sections for $\eta$ photoproduction off protons have been 
measured at $E_\gamma$ = 1.6 -- 2.4 GeV in the backward direction. 
A bump structure has been observed above 2.0 GeV in the total energy. 
No such bump is observed in $\eta', \omega$ and $\pi^0$ photoproductions. 
It is inferred that this unique structure in $\eta$ photoproduction is 
due to a baryon resonance with a large $s\bar{s}$ component which is strongly 
coupled to the $\eta N$ channel. 

\end{abstract}


\maketitle


The constituent quark model has been very successful 
in describing the ground state of the flavor SU(3) octet and decuplet baryons. 
Many baryon resonances predicted by the constituent quark model have been 
discovered and are well established~\cite{PDG,Isgur,Capstick}. 
There are two well-known problems for the 3-quark model of baryons. 
One is the mass-order-reverse problem for the lowest excited 
state~\cite{N1535}.  
According to the constituent quark model, the negative-parity state with 
the orbital angular momentum, $L$ = 1, should be the lowest except for proton 
and neutron. 
However, the lowest has been experimentally found as $S_{11}$(1535), which is 
heavier than $P_{11}$(1440), and $\Lambda$(1405) with a $s$-quark. 
This mass-order-reverse problem may be solved by taking into account the extra quark and 
anti-quark pair. 
A large admixture of $s\bar{s}$ for $S_{11}$(1535), $u\bar{u}$ for 
$P_{11}$(1440) and $d\bar{d}$ for $\Lambda$(1405) was proposed to explain 
the mass order of these resonances~\cite{N1535}. 
As a natural consequence of a large $s\bar{s}$ admixture in $S_{11}$(1535), 
we expect a strong coupling to $\eta N$ because $\eta$ meson is the 
lightest meson with a $s\bar{s}$ component. 

The other problem is concerned with so-called ``missing baryon resonances''. 
Many resonances around 2 GeV are predicted in the constituent quark model. 
However, a large number of them are not identified experimentally. 
Information on baryon resonances mainly comes from the pion induced 
productions. A part of missing resonances may couple to 
other mesons such as $\eta N$, $\omega N$, $K\Lambda$, 
etc, and could escape from search~\cite{Capstick}. 
Study of $\eta$ photoproduction has the advantage of looking for specific resonances 
with a large $s\bar{s}$ component and an isospin of $\frac{1}{2}$ by means of decay into the 
$\eta$N. 

Recently, $\eta$ photoproduction has been studied by the CB-ELSA and CLAS 
collaborations. 
An enhancement of differential cross sections was observed around 
$W$ = 1.85 GeV in the total energy,  
and was claimed to be due to the third $S_{11}$ resonance by 
CLAS~\cite{eta-jlab}. 
On the other hand, CB-ELSA found evidence of a new resonance $D_{15}$(2070), 
but did not observe the third $S_{11}$ resonance~\cite{eta-bonn}. 
The BES collaboration found a peak around 2065 MeV in the invariant 
mass spectrum of $\pi N$ from $J/\psi$ decay into $\bar{N}N\pi$~\cite{BES2}. 
The peak is suggested to be due to the $P_{11}$(2100) resonance, which is 
predicted to strongly couple to $\eta N$ by the Pitt-ANL 
model~\cite{PDG,Vrana}. 
The $\pi N$ system in this decay mode has an isospin $\frac{1}{2}$. 
All the experimental results indicate that there are some unestablished 
resonances around 2 GeV. 
Therefore, it is expected that the additional precise 
measurement of $\eta$ photoproduction sheds light to clarify the nature of 
these resonances. 

In this letter, we report the differential cross sections of backward-angle 
$\eta$ photoproductions from protons in the energy range $E_{\gamma} = 1.6  
- 2.4$ GeV by detecting protons at 
forward angles to identify $\eta$ mesons in the missing-mass spectrum. 
At backward angles, no appreciable contribution is expected from forward 
diffractive processes, and  it is expected that 
one can clearly observe resonances. 

The experiment was carried out at the SPring-8/LEPS facility~\cite{LEPS}. 
The laser-electron photon beam was produced by backward-Compton scattering 
between Ar-ion laser photons with a 351-nm wave length and electrons with 
the 8 GeV energy. The photon energy range was from 1.6~GeV to 2.4~GeV. 
The energy resolution was 10 MeV in root-mean-square. 
A liquid hydrogen target with a thickness of 16.5 cm was used.  
The data were accumulated with 1.0$\times 10^{12}$ photons at the 
target. Charged particles were detected by using the LEPS magnetic spectrometer. 
The angular coverage of the spectrometer was 
about $\pm 20^\circ$ and $\pm 10^\circ$ in the horizontal and vertical 
directions, respectively. 
Charged particle events with more than 98\% of confidence level of a track fitting 
were used in this analysis. 
Mass identification was made using the momentum, the path length, and 
the time-of-flight. 
The momentum range of protons was 1.4 -- 2.4 GeV/c in the present analysis. 
The momentum resolution for 1.4 -- 2.4 GeV/c  
protons was 0.7 -- 0.9\%. 
The proton mass resolution was 46 MeV/c$^2$ at 2~GeV/c momentum. 
Proton events were selected in the reconstructed mass spectrum within 
4$\sigma$ of the nominal value~\cite{pi0}. 
Contaminations from pions and kaons were estimated to be less than 0.1\%. 
Reaction vertex points were reconstructed as the closest point between a 
track and the beam axis, and were used to select events produced from the liquid 
hydrogen target. Some contaminations came from the events produced in 
a charge-defining plastic scintillator placed behind the target. 
The contamination rate was typically 1\%, but increased up to 5\% 
at a scattering angle less than six degrees because of poorer 
vertex reconstruction resolution. 
The spectrometer acceptance, including the efficiency for detectors and 
track reconstruction, was obtained using a Monte Carlo simulation with 
the GEANT3 code~\cite{geant}. The acceptance, which depended on the photon 
energy and the scattering angle, was calculated.    
Detail descriptions about the detectors and particle identification are given 
in Ref.~\cite{LEPS,pi0}. 

\begin{figure}[h]
\includegraphics[height=8cm]{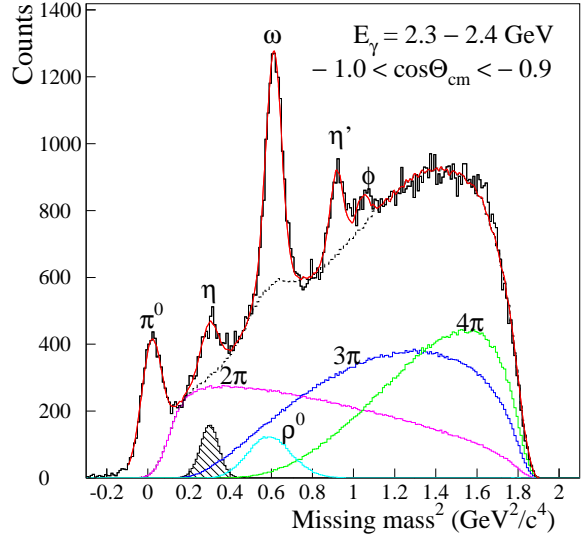}
\caption{(Color online). Spectrum of missing mass square for $\gamma p 
\rightarrow p X$ at $E_\gamma$ = 2.3 -- 2.4 GeV and cos$\Theta_{c.m.}$ 
= --1.0 $\sim$ --0.9, where $\Theta_{c.m.}$ is a scattering angle of mesons. 
The peaks are due to $\pi^0$, $\eta$, $\omega$, $\eta'$ and $\phi$ 
meson photoproductions. The red curve is the fitting results. 
The pink, blue, green and light-blue curves are for 2$\pi$, 3$\pi$, 
4$\pi$ and $\rho^0$ photoproductions determined by a fit to the data, 
respectively.  The dashed curve is the distribution of the sum of them. 
The hatched histogram is for $\eta$ meson. 
\label{MM}}
\end{figure}
The data were divided into eight energy bins from 1.6 to 2.4 GeV, and 4 
angular bins from --1.0 to --0.6 in cos$\Theta_{c.m.}$, where $\Theta_{c.m.}$ 
is a scattering angle of mesons in the center of mass system. 
The spectrum of missing mass square for the $\gamma p \rightarrow p X$ 
reaction is shown in Fig.~\ref{MM}. 
The peaks observed are due to $\pi^0$, $\eta$, $\omega$, $\eta'$ and $\phi$ 
meson photoproductions. 
The background under the $\eta$ meson peak consists of non-resonant 
2$\pi$, 3$\pi$, and $\rho^0$ photoproductions. 
The missing mass distributions for each photoproduction at different energies 
and at different angular bins were generated in the Monte Carlo simulation by 
taking into account the detector resolution of the present experiment. 
The experimental data were fitted by summing all the generated missing-mass 
distributions of signal and background processes. 
The contribution of each reaction channel were determined by the 
relative height of the distribution  in order to minimize the fitting $\chi^2$. 
The reduced $\chi^2$ was 1.2 at minimum, and 2.3 at maximum. 
The contributions of each reaction channel in the fitting are shown 
in Fig.~\ref{MM}. 
Finally, the yield of $\eta$ was extracted from the fit. 
The energy dependence of cross sections for the non-resonant 
multi-pion productions was estimated from the fitting, and 
assumed to be smooth in determining the generated distributions. 
The distribution for $2\pi$ production, which is the main background for 
$\eta$ photoproduction, gradually decreases toward high energies.  

Experimentally, the statistics of $\eta$ photoproduction increased with 
the scattering angles of $\eta$ meson and with energies since  
the spectrometer acceptance was larger at smaller angles of protons, and 
the number of photons is larger at higher energies. 
The statistical error is 4\% at minimum at $W$ = 2.20 -- 2.24 GeV and at 
--1.0 $<$ cos$\Theta_{c.m.} <$ --0.9, and is 13\% at maximum at $W$ = 1.97 
-- 2.02 GeV and at --0.7 $<$ cos$\Theta_{c.m.} <$ --0.6. 
Background events from the charge-defining plastic scintillator behind 
the target make a broad missing mass distribution due to the Fermi motion. 
The contamination rate was estimated to be 3\% and 0.5\% 
at --1.0 $<$ cos$\Theta_{c.m.} <$ --0.9 and --0.9 $<$ cos$\Theta_{c.m.} 
<$ --0.6, respectively, and was subtracted from the yields. 

The systematic uncertainty for the resolution estimated in Monte Carlo 
simulation is obtained from the $\eta$ peak fitting by using Gaussian 
function. 
The systematic uncertainty is 4\% at minimum, and 20\% at maximum. 
The systematic uncertainty for the background subtraction due to the 
uncertainty of the energy dependence was obtained to be 0.1 -- 3.3\%. 
The systematic uncertainty for the target thickness, due to fluctuations 
of the temperature and pressure of the liquid hydrogen, was estimated to be 
1.0\%. The systematic error of the photon number normalization was 3.0\%. 
The systematic uncertainty of the aerogel \v{C}erenkov counter (AC) due to 
accidental vetoes and $\delta$-rays was measured to be lower than 
1.6\%. 
The overall systematic uncertainties were calculated as the square root 
of the quadratic sum of these systematic uncertainties, and were 3.6 -- 
20.1\%.  


\begin{figure}[h]
\includegraphics[height=8cm]{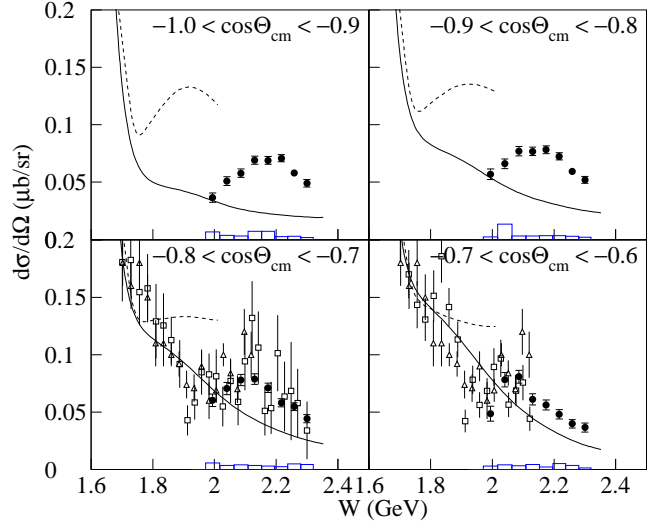}
\caption{(Color online). Differential cross sections for $\eta$ photoproduction 
plotted as a function of the total energy, $W$. The angular range is indicated.  
The closed circles are the present results.
Statistical error is plotted as error bars, and the blue 
boxes show systematic uncertainties.  
The open triangles and open squares are the experimental data from 
CLAS~\cite{eta-jlab} and CB-ELSA~\cite{eta-bonn}, respectively. 
The solid and dashed curves are the results of SAID~\cite{SAID} and
ETA-MAID~\cite{MAID}, respectively. 
\label{eta}}
\end{figure}
Figure~\ref{eta} shows the differential cross sections for $\eta$ 
photoproduction. 
The differential cross sections have been obtained for the first time 
in the angular range of $-$1 $<$ cos$\Theta_{c.m.} <$ $-$0.8. 
A wide bump structure has been observed above 2.0 GeV in the total energy, 
$W$. 
The present results are consistent with the data from CLAS and CB-ELSA that 
indicate the presence of a bump structure around $W$ = 2.1 GeV. 

The maximum cross section of the bump amounts to about 0.08 $\mu b$/sr on top 
of the continuum background, which may come from the non-resonant process.  
The background decreases with increasing energy (shown by the solid curves 
in Fig.~\ref{eta}). Therefore, the bump is comparable to the background. 
In order to explain the presence of a large bump, a contribution from 
resonances is required. 
The central position of the bump structure shifts to higher energies at 
backward angles. The locations are $W \sim$ 2.06 GeV at --0.7 $<$ 
cos$\Theta_{c.m.} <$ --0.6, and $W \sim$ 2.17 GeV at --1.0 $<$ 
cos$\Theta_{c.m.} < $ --0.9. 
This is possible because an interference between resonances and 
diffractive processes may depend on the scattering angles.  
The other possibility is that the bump structure consists of more than one resonance, 
whose angular distributions are different.  

The SAID calculations are consistent with the experimental data below 2.0 GeV, 
but underestimate the data above 2 GeV.  An additional process 
is needed for explaining the bump structure above 2 GeV.  
In the SAID analysis, the possible resonance is $G_{17}$(2190) with 
($M,\Gamma$) = (2152, 484) MeV around 2.1 GeV~\cite{igor}. 
The results of ETA$-$MAID are consistent with the data below 1.8 GeV.  
However, the bump structure does not appear in the calculation~\cite{MAID}. 
Although the central position of the bump structure in the present 
analysis moves from 2.06 to 2.17 GeV, the position and the bump width 
is roughly consistent with those of the resonance $D_{15}$ with (M, $\Gamma$) 
= (2068 MeV, 295 MeV) as suggested by the CB-ELSA collaboration 
work~\cite{eta-bonn}. 
Detailed studies including the precise angular distribution at backward 
angles will help to improve theoretical calculations in revealing the hidden 
resonance states contributing to the bump structure. 

\begin{figure}[h]
\includegraphics[height=8cm]{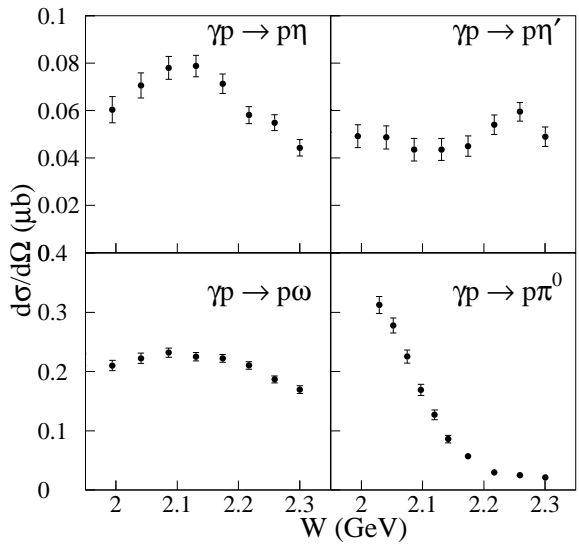}
\caption{Differential cross sections for $\eta, \eta', \omega$ and $\pi^0$ 
photoproductions plotted as a function of the total energy ($W$) at 
$-0.8 < $cos$\Theta_{c.m.} < $ --0.7. 
\label{etap-omega}}
\end{figure}
Figure~\ref{etap-omega} shows the differential cross sections for $\eta$, 
$\eta'$, $\omega$ and $\pi^0$ photoproductions. 
Differential cross sections for $\eta'$, $\omega$ and $\pi^0$ 
photoproductions have been simultaneously obtained in the present analysis. 
The results on $\pi^0$ photoproduction have been reported in Ref.~\cite{pi0}.  
A small bump structure is observed around 2.25 GeV in $\eta'$ photoproduction. 
There is no prominent structure for the $\omega$ photoproduction 
cross-section.   
The cross section for $\pi^0$ photoproduction drastically decreases with 
energies up to 2.15 GeV and shows a mostly flat distribution above 2.15 GeV. 
The energy dependence of cross sections for $\pi^0$, $\eta'$ and $\omega$ 
photoproductions are different from that for $\eta$ photoproduction. 
The wide bump structure is a specific character of $\eta$ photoproduction. 
Therefore, the bump structure is expected to be due to resonances with a 
 large $s\bar{s}$ component and with a strong coupling to the $\eta N$ 
channel. 

In summary, $\eta$ photoproduction off protons has been measured at  
$E_\gamma$ = 1.6 -- 2.4~GeV at the SPring-8/LEPS facility. 
The differential cross sections have been obtained with high statistics at 
backward scattering angles of $\eta$ mesons in the $\gamma p 
\rightarrow p \eta$ reaction by detecting 
protons scattered at forward angles. 
This work provides unambiguous evidence for a bump structure above $W$ = 2.0 GeV. 
No such structure is  
seen in $\eta'$, $\omega$, and $\pi^0$ photoproductions. 
It is inferred that this unique structure in $\eta$ photoproduction is due to 
a baryon resonance with a large $s\bar{s}$ component which strongly coupled 
to the $\eta N$ channel. 


\begin{acknowledgments}
We thank the staff at SPring-8 for providing excellent experimental 
conditions. We thank I.~I.~Strakovsky for fruitful discussions on the possible 
resonances around 2.1 GeV.  
This work was supported in part by the Ministry of Education, 
Science, Sports and Culture of Japan, by the National Science Council 
of the Republic of China (Taiwan), and by the National Science Foundation 
(USA).
\end{acknowledgments}


\begin{thebibliography}{99}

\bibitem{PDG} Particle Data Group, Phys.~Lett.~B {\bf 667}, 1 (2008). 
\bibitem{Isgur} N.~Isgur and G.~Karl, Phys.~Rev.~D {\bf 18}, 4187 (1978); 
N.~Isgur and G.~Karl, Phys.~Rev.~D {\bf 19}, 2653 (1979). 
\bibitem{Capstick} S.~Capstick and W.~Roberts, Phys.~Rev.~D {\bf 49}, 4570 (1994); 
S.~Capstick and W.~Roberts, Phys.~Rev.~D {\bf 58}, 074011 (1998);
S.~Capstick and W.~Roberts, Prog.~Part.~Nucl.~Phys. {\bf 45}, S241 (2000). 
\bibitem{N1535} B.~S.~Zou, Nucl.~Phys.~A {\bf 790}, 110c (2007). 
\bibitem{eta-jlab} M.~Dugger {\it et al.} (CLAS Collaboration), Phys.~Rev.~Lett. {\bf 89}, 222002 (2002).
\bibitem{eta-bonn} V.~Crede {\it et al.} (CB--ELSA Collaboration), Phys.~Rev.~Lett. {\bf 94}, 012004 (2005).
\bibitem{BES2} M.~Ablikim {\it et al.} (BES Collaboration),  
Phys.~Rev.~Lett. {\bf 97} 062001 (2006).
\bibitem{Vrana} T.~P.~Vrana, S.~A.~Dytman, and T.~-S.H.~Lee, Phys.~Rep. {\bf 328}, 181 (2000). 
\bibitem{LEPS} M. Sumihama {\it et al.} (LEPS Collaboration), Phys. Rev. C {\bf 73}, 035214 (2006); T.~Nakano {\it et al.} (LEPS Collaboration), Phys. Rev. C {\bf 79}, 025210 (2009).
\bibitem{pi0} M.~Sumihama {\it et al.} (LEPS Collaboration), Phys.~Lett.~B {\bf 657}, 32 (2007).
\bibitem{geant}R.~Brun {\it et al.}, Applications Software Group, CERN Program 
Library Long Writeup W5013. 
\bibitem{SAID} R.~Arndt, W.~J.~Briscoe, I.~I.~Strakovsky, and R.~L.~Workman,
Phys.~Rev.~C {\bf 74}, 045205 (2006); http://gwdac.phys.gwu.edu/.
\bibitem{MAID}L.~Tiator and S.~Kamalov, nucl-th/0603012;
D.~Drechsel, O.~Hanstein, S.S.~Kamalov and L.~Tiator,
Nucl.~Phys.~A {\bf 645}, 145 (1999); http://www.kph.uni-mainz.de/MAID/.
\bibitem{igor} I.~I.~Strakovsky (private communication). 

\end{thebibliography}

\end{document}